# Free energy along drug-protein binding pathways interactively sampled in virtual reality

Helen M. Deeks,[1,†] Kirill Zinovjev,[2,3†] Jonathan Barnoud,[1,4] Adrian J. Mulholland,[1] Marc W. van der Kamp,[1,3*] David R. Glowacki,[4*]

[†]*These authors contributed equally this work*

[1]*Center for Computational Chemistry, School of Chemistry, University of Bristol, Bristol, BS8 1TS, UK,*
[2]*Departamento de Química Física, Universidad de Valencia, Burjassot, 46100, Spain,* [3]*School of Biochemistry, University of Bristol, Bristol, BS8 1TD, UK,* [4]*CiTIUS Intelligent Technologies Research Centre, Rúa de Jenaro de la Fuente, s/n, 15705 Santiago de Compostela, A Coruña, Spain*

*\*marc.vanderkamp@bristol.ac.uk; \*drglowacki@gmail.com*

## Abstract

We describe a two-step approach for combining interactive molecular dynamics in virtual reality (iMD-VR) with free energy (FE) calculation to explore the dynamics of biological processes at the molecular level. We refer to this combined approach as iMD-VR-FE. Stage one involves using a state-of-the-art iMD-VR framework to generate a diverse range of protein-ligand unbinding pathways, benefitting from the sophistication of human spatial and chemical intuition. Stage two involves using the iMD-VR-sampled pathways as initial guesses for defining a path-based reaction coordinate from which we can obtain a corresponding free energy profile using FE methods. To investigate the performance of the method, we apply iMD-VR-FE to investigate the unbinding of a benzamidine ligand from a trypsin protein. The binding free energy calculated using iMD-VR-FE is similar for each pathway, indicating internal consistency. Moreover, the resulting free energy profiles can distinguish energetic differences between pathways corresponding to various protein-ligand conformations (e.g., helping to identify pathways that are more favourable) and enable identification of metastable states along the pathways. The two-step iMD-VR-FE approach offers an intuitive way for researchers to test hypotheses for candidate pathways in biomolecular systems, quickly obtaining both qualitative and quantitative insight.

# Significance statement

Drug activity often depends on how long they stay bound to their targets, i.e. the lifetime of the complex, which is determined by the process of dissociation. Finding unbinding pathways, and the energy barriers along them, is increasingly important in drug development. Simulating these slow processes requires enhanced sampling molecular dynamics methods, and often a guess of the path. Human spatial intuition can navigate such hyperdimensional configuration spaces, but is difficult to express as instructions to a simulation algorithm. Here, we show that interactive simulations in VR – enabling users to 'reach out and touch' the moving molecules – can guide a ligand along physically reasonable pathways out of a protein. The probability of pathways is then quantified using free energy calculations.

# 1. Introduction

Recent advances in virtual reality (VR) technology have enabled new workflows across several scientific and engineering domains. For example, recent applications in nanoscience and microscopy use VR for interactively manipulating real-time dynamics of physical systems, guided by human scientific insight. (1-3) For understanding (bio)molecules, immersive technologies like VR have significant potential, given the fact that many molecular systems are characterized by considerable 3d complexity. The majority of work applying VR to molecular systems tends to focus on *interactive visualization*, e.g., refs (4-6). Over the last few years, we have published a number of studies outlining strategies for *interactive simulation*, using an approach which we call interactive molecular dynamics in virtual reality (iMD-VR). Narupa, our open-source iMD-VR framework (7), enables users to interact with a real-time molecular simulation as if it were a tangible dynamic object. Within this virtual environment, the player can reach out with a 'force probe' (i.e. a VR controller) and interactively manipulate the dynamics of molecular motion. In this way, a researcher can use their insight and expertise to guide molecules, using both spatial and chemical intuition to explore states of interest. (8)

The ability to manipulate molecules as if they were tangible objects is a unique means of studying molecular transformations, mechanisms, and rare events. Previous work has demonstrated that iMD-VR has acceleration benefits (2 – 10x) for performing 3d molecular tasks compared to 2d interfaces. (7, 9) We have also shown iMD-VR to be a useful tool in building protein-ligand complexes (10), with recent application to the SARS-CoV-2 main protease. (11, 12) By coupling iMD-VR to quantum mechanical methods, we have also shown that reaction pathways can be efficiently sampled using iMD-VR. (13) When machine learning algorithms (e.g., atomic neural networks) are trained on these human-sampled reactive pathways, the learning rate is nearly 10x faster than data sets obtained through more conventional brute-force sampling approaches. (14, 15) Here, we show how iMD-VR can be used practically as a tool for enhancing the sampling of pathways through biomolecular conformational space, providing input for efficient free energy calculations that in turn provide the iMD-VR user with feedback.

Protein-ligand systems are high-dimensional and continually fluctuate between different conformations, often separated by kinetic or thermodynamic barriers. Such systems are increasingly being studied using molecular dynamics (MD) simulations, providing insight into their behaviour (16, 17) However, the dynamic and structural complexity of protein-ligand complexes makes them challenging to simulate. A single ligand unbinding event can take milliseconds, or even seconds, to occur. (18, 19) While sampling of rare events in equilibrium simulations is computationally demanding and often not feasible, accelerated sampling techniques, such as umbrella sampling (20) and metadynamics (21), can be employed to reduce this load. However, many of these methods require defining a reaction coordinate (RC) along which to bias the simulation. A simple example of an RC would be to describe bond breaking as a single interatomic distance; here, the RC is made up of a single collective variable (CV), i.e., the bond distance. When applied to more complex molecular transitions however, basic RCs may not encode all the motions relevant to a molecular process (and therefore would not accurately control the progress of the transition). Defining an RC that uses more 'collective variables' (CVs) enables more sophisticated biasing, however, care still needs to be taken: A more detailed RC does not guarantee better guidance along the minimum free energy path of the transition (and may even be deleterious if CVs are included that are not relevant to the progress of the transition). (22)

To describe complex molecular processes that can be understood as transitions between (meta)stable states, path collective variables (pathCVs) are one approach for defining RCs in high-dimensional space. Realistic trajectories of these processes (e.g. protein-ligand binding) are expected to be close to the minimum free energy path (MFEP) connecting those states. In such cases, a RC for the process can be defined as a path collective variable (pathCV) that changes smoothly when the system advances along the MFEP. Although a pathCV can be defined using MD snapshots along the MFEP, more sophisticated approaches will apply a transformation to this data, for example, by creating internal coordinates, (23, 24) assigning different weights to the CVs in the distance calculation, (25) or employing the metric tensor defined by the geometry of the CV space. (26) Application of pathCVs in enhanced sampling can enable the sampling of complex processes, including enzyme-catalyzed reactions (27) and large scale conformational changes in proteins. (28)

Sampling a process in a molecular system using pathCVs can be broken down into two stages: (i) discovering an (approximate) MFEP that connects two states and (ii) sampling along a pathCV defined using this MFEP. While the latter can be done with any enhanced sampling technique, it relies on being able to resolve a MFEP in the first place; a challenge in itself. Interactive molecular simulation environments (such as iMD-VR) can be used to take advantage of human spatial and chemical intuition of a complex conformational landscape. With careful direction using iMD-VR, a ligand can be placed into, or removed from, a protein binding pocket in 10-100 picoseconds of simulation time (taking only minutes of actual time). (10) Tests of several protein-ligand systems showed that users (including non-specialists) can generate structures similar to those obtained by protein crystallography. Similarly, iMD-VR can be employed to generate candidate (un)binding pathways that explore an ensemble of conformations.

In this work, we show how iMD-VR and free energy sampling techniques can be effectively combined to aid exploration of high-dimensional biomolecular systems, using as an example ligand dissociation from proteins. Fig. 1 illustrates the workflow proposed here. Seven unbinding trajectories, or human-sampled paths, were generated within an iMD-VR simulation of the trypsin-benzamidine complex, a protein-ligand system with well characterized energetics. (29-32) Following ref. (33), we then projected snapshots along these paths into the space of six CVs that capture the position and orientation of benzamidine relative to trypsin. Using this reduced-dimensional descriptor, the free energies along these initial 'guess' pathways were calculated. We also explore how the adaptive string method (34) can be used to gain quantitative suggestions for how inputs from iMD-VR can be optimized. The iMD-VR simulation presented here is available as a cloud-hosted iMD-VR service. Instructions on how to connect to the cloud simulation are given in the SI Appendix, alongside input files for running the iMD-VR simulation locally.

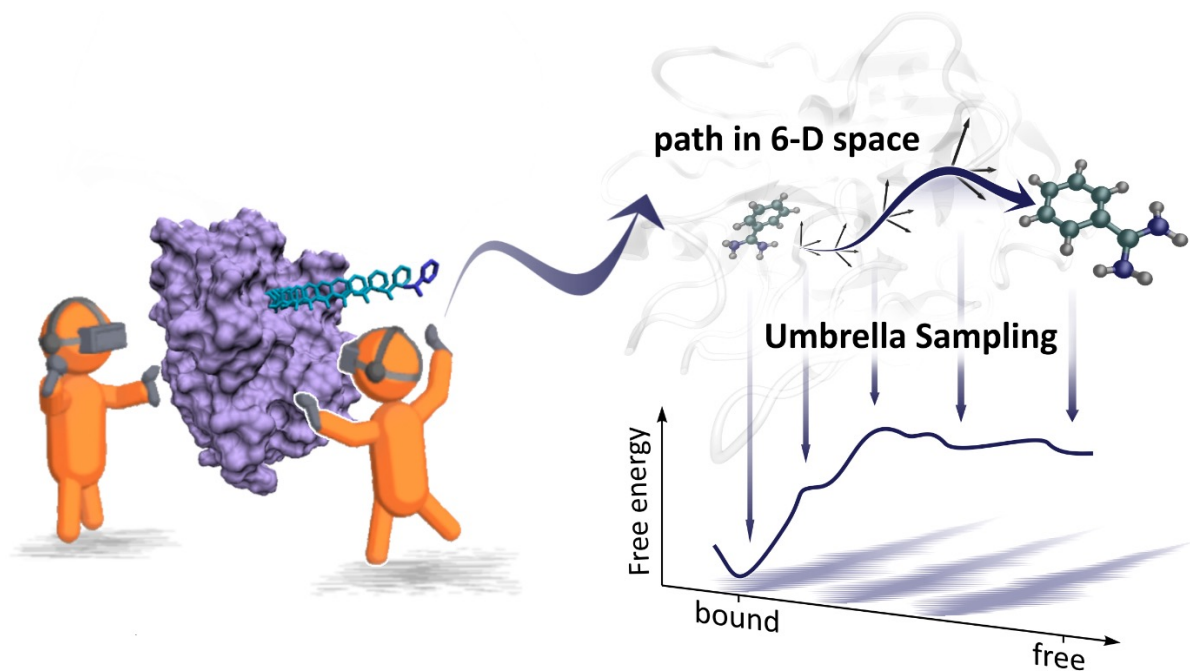

**Figure 1.** Workflow used to obtain free energy profiles for benzamidine unbinding from trypsin. First, users in iMD-VR model the dissociation of the ligand, by 'pulling' it out in an interactive MD simulation (left). Snapshots of the unbinding trajectory are used as input for umbrella sampling along a 6-D path collective variable (right) to obtain the free energy profile along the path.

## 2. Results

### 2a. Generation of iMD-VR pathways

Seven of the iMD-VR generated benzamidine unbinding pathways are shown in Fig. 2, each overlaid over a static representation of the trypsin surface. Starting from the bound complex, the user applied forces to the benzamidine ligand by selecting specific atoms and 'pulling' the controller in the desired direction, in order to move the ligand around relative to the protein. These forces are included alongside the regular force-field forces in the iMD simulation. Altogether, these human-sampled (H-S) paths were generated within an hour of laboratory time. Aside from an indication of the protein surface (through rendering the Van der Waals radius of each atom), the simulations did not include any specific visual guidance; the user simply aimed to move the ligand away from the protein. The substrate binding pocket (denoted as S1) sits buried in a larger groove on the protein surface. Restraints on the protein backbone were used to avoid large protein conformational changes (see methods). Within the 3D iMD-VR environment, the researcher hypothesized multiple unbinding paths. H-S path 1 (red in Fig. 2) does not explore surface interactions; instead, benzamidine is pulled directly towards the bulk space. H-S paths 2-4 (orange, yellow, and green in Fig. 2) directed benzamidine away from the His57-Asp102-Ser195 catalytic triad and explored the steeper sides of the substrate pocket. In contrast, in H-S paths 5-7 (light blue, dark blue, and purple in Fig. 2) guided benzamidine was guided through the substrate binding pocket, moving it past the catalytic triad. Beyond that point, the binding pocket splits in two grooves extending in different directions. H-S path 5 did not explore this bifurcation and instead guided benzamidine straight into the bulk. However, for H-S paths 6 and 7, the researcher moved benzamidine down along one or other of these two grooves. iMD-VR provides a convenient approach to explore alternative pathways in 3D space. (9) Here, upon seeing a clear groove in the protein surface during the sampling of H-S path 7, the researcher decided to guide benzamidine close to the protein surface.

## 2b. Free energy sampling

Fig. 2 shows the free energy along each of the seven iMD-VR trajectories (or H-S paths). The reaction coordinate was defined as a pathCV in a space of 6 CVs describing the ligand position and orientation relative to the protein (as proposed in ref. (33); see SI Appendix for details). Free energy profiles were obtained using umbrella sampling (US) MD simulations in two regimes: 2a shows the profiles integrated using only 10 ps of sampling in each of the 56 US windows, while 2b shows the profiles obtained from 1 ns sampling per window. The profiles from longer US are smooth, with differences in unbinding free energies within a few kcal/mol, indicating good convergence.

Not surprisingly, the free energy profiles obtained from short sampling (Fig. 2a) are much noisier and show significant variation of the estimated unbinding free energy. However, the most prominent features of the profiles obtained from longer sampling (approximate heights of the barriers and positions of the intermediate states) are already apparent. Thus, such ultra-short US provides quick estimate of the free energy along the ligand dissociation paths obtained from iMD-VR.

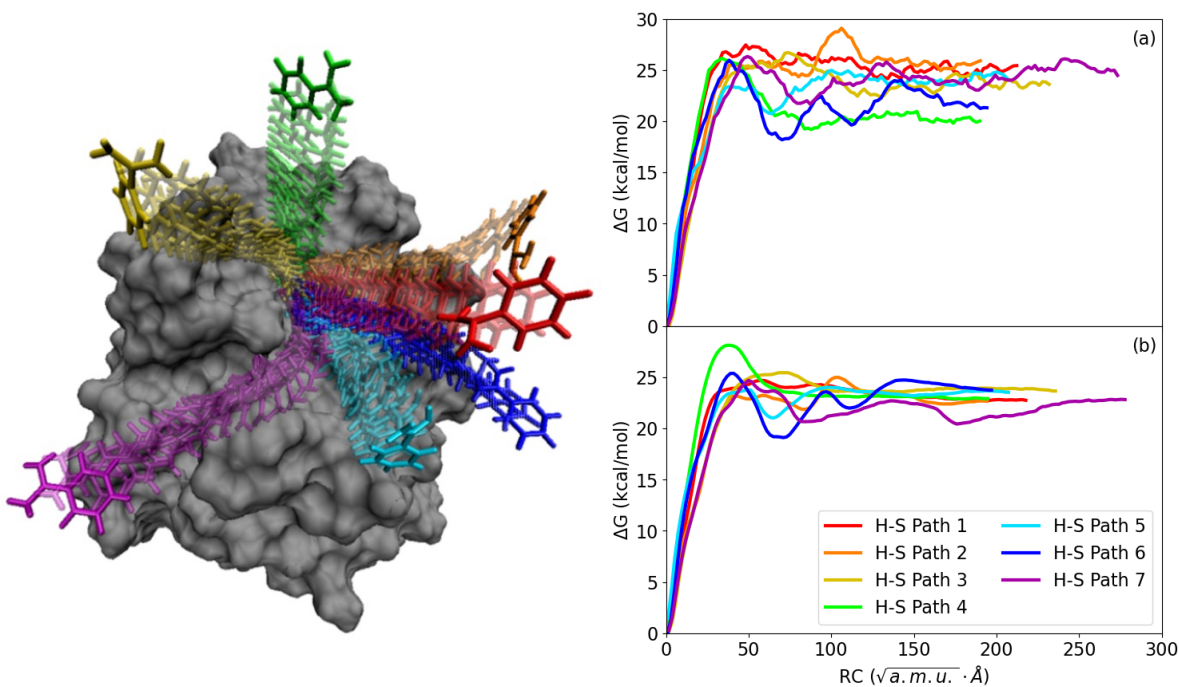

**Figure 2.** Benzamidine unbinding pathways and their free energy profiles. Left: The seven human-sampled benzamidine unbinding pathways obtained from iMD-VR superimposed onto the trypsin starting structure (based on PDB ID 1S0R). The color of each route corresponds to the line colors in the right-hand panel. Right: Free energy profiles calculated using the weighted histogram analysis method from 56 umbrella sampling windows along each of the 7 human-sampled paths. (a) Profiles obtained using only 10 ps of sampling in each US window; (b) profiles obtained using 1 ns of sampling per window.

## 2c. Energetic characterization of iMD-VR generated pathways

The free energy of binding was estimated to be −22.5 kcal/mol, larger than the experimental value of −6.2 kcal/mol. (35) Our simulation protocols used an implicit solvation model (that underestimates the benzamidine solvation energy), alongside positional restraints on the protein (that stabilize the bound state). Therefore, a significant overestimation of the binding free energy is expected. Nonetheless, between paths, the estimated (un)binding free energy (the difference in energy between the starting complex and final unbound state) remained within a few kcal/mol, indicating that the obtained free energy profiles are well converged.

Most paths had a barrier height of approximately 25 kcal/mol corresponding to the iMD-VR user breaking the electrostatic contact with Asp189. However, H-S path 4 had a higher barrier as benzamidine was guided in a perpendicular direction to the S1 pocket opening, resulting in steric clashing against the roof of the pocket. H-S paths 1-4, which guided benzamidine away from the catalytic triad, generally had larger barrier heights. Of these four, path 2 guided benzamidine into a hydrophobic basin surrounding the S1 pocket, where the trypsin surface residues appear to adapt and form a cavity around the ligand. However, the user inadvertently guided the benzyl group against the polar Ser96 and Asp97 surface residues and caused a spike in free energy. The small minimum at a similar point for H-S path 6 is caused by the benzamidine being oriented such that the charged amidine group runs past these residues instead. Comparing the data in this work to the metastable states identified in ref. (30), H-S path 2 moved the ligand closest to state S1, H-S path 6 moved the ligand closest to state S3, and H-S path 7 moved the ligand closest to state S2.

Fig. 3 shows the free energy profile of path 7 after refinement by ASM of the initial iMD-VR generated (H-S) path. There is a marked decrease in the free energy as benzamidine exits the S1 pocket. Specifically, two intermediate states are formed, corresponding to benzamidine rotating itself out of the S1 pocket. Figs. 3a-c show representative snapshots of these states. In the first state, benzamidine rotates so that the contact with Asp189 is broken, but the hydrophobic group is buried in the space just outside the S1 pocket and favourable interactions are formed between the backbone oxygens of Gly214 and the Gln192 residue. In the second state, benzamidine has fully rotated itself such that the benzyl group is buried in the hydrophobic basin, specifically sandwiched between the alpha carbons of Cys191 and Trp211. The polar, charged amidine group is pointed towards the solvent and forms a closer interaction with Gln192. Notably, these two intermediate states are similar to states B and P described by Tiwary *et al.*, even though the reported water-mediated interactions are not captured due to our use of implicit solvent. (36) Nonetheless, the ASM refinement identifies new, stable states that are approximately 7 kcal/mol lower in energy than those sampled using iMD-VR alone, giving quantitative feedback on how future iMD-VR sampling could be improved (by guiding the ligand through such stable states). After these intermediate states, benzamidine remains close to the original iMD-VR path.

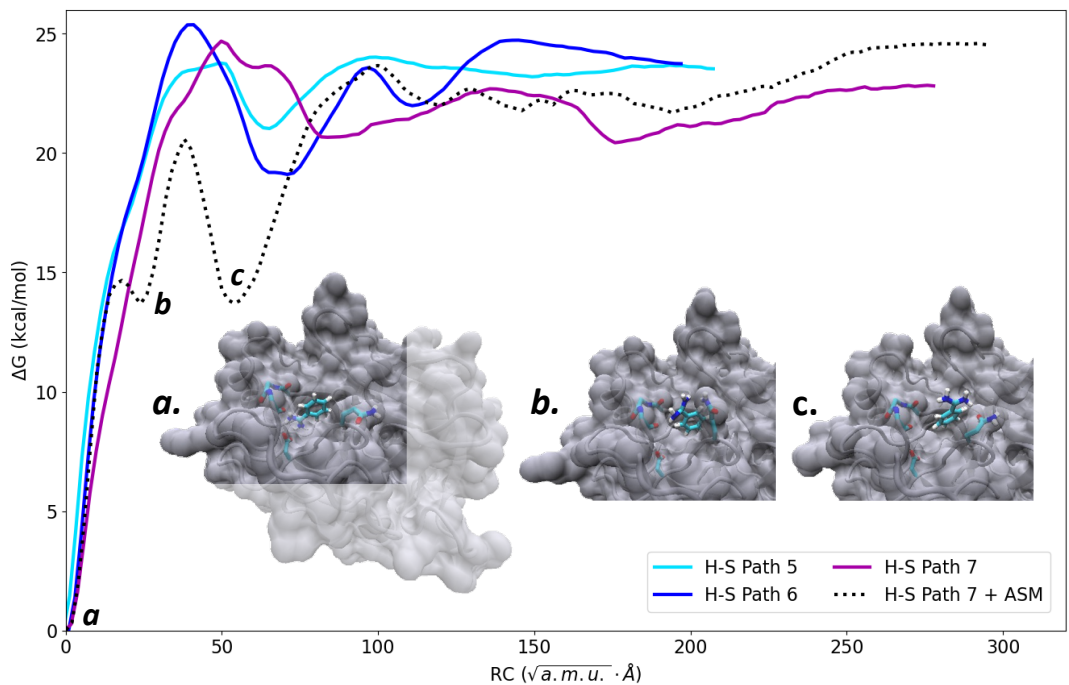

**Figure 3.** Free energy profile for path 7 refined using the adaptive string method (ASM). The dotted line and snapshots of key transitions depict the ASM-refined pathway, with the continuous lines for human-sampled paths 5-7 (from Fig. 2) shown for comparison. Representative snapshots from three points along the ASM-refined free energy profile are overlaid, showing how benzamidine rotates out of the binding pocket: (a) the starting bound state, (b) the first intermediate state, and (c) the second intermediate state.

## 3. Discussion

iMD-VR is an emerging tool for the quick exploration of complex molecular environments. Within a single hour-long laboratory session, iMD-VR was used here to generate seven unbinding pathways for benzamidine exiting the trypsin S1 pocket. Each H-S path served as a 'guess' for how unbinding could happen, driven by the iMD-VR user's own chemical and spatial intuition, alongside the forces from molecular dynamics. With reference to just a single distance, two angles and three dihedral angles from each molecular snapshot, any protein-ligand iMD-VR trajectory can be projected into a six-dimensional pathCV (see methods and SI Appendix for details). As a result, inputs from iMD-VR can be used as the (initial) bias in enhanced sampling.

A carefully chosen RC can significantly reduce the computational resources required to simulate unbinding processes. However, selecting the CVs that capture the movement of a small ligand relative to a large, constantly fluctuating protein is a difficult challenge. (22, 37) With increasing system complexity, hand selecting descriptors that are both comprehensive and high quality is impractical, especially where multiple pathways are being considered, as is often the case. Our aim was to perform a quick comparison of the free energy profiles related to the iMD-VR trajectories, and so we factored in how much sampling was needed for convergence. While a simple RC (such as distance between the ligand and some group in the active site) would be fast to implement, it would not contain any information about the direction and rotation of the ligand. Therefore, there is little guarantee that the sampling would follow the iMD-VR trajectory. On the other hand, an arbitrarily complex RC will probably better follow the iMD-VR paths, but may also include redundant information (such as conformational changes in the protein or ligand unrelated to the path, sampled accidentally in iMD-VR) without adding much value. We identified a set of six positional and rotational descriptors (see SI Appendix) as the minimal amount of information needed to unambiguously identify the ligand position and orientation with respect to the protein. This approach is particularly suitable for representing iMD-VR pathways because these descriptors can be measured during the interactive simulations, immediately projecting the user-sampled pathway onto a reduced dimensional space.

We demonstrate that 10 ps US per window is sufficient to provide a reasonable approximation of the underlying free energy profile (fig. 2). The protocol demonstrated here can be expanded to 'on-the-fly' integration with VR, which could eliminate *a posteriori* analysis of the full iMD-VR pathways and help guide the user. For example, H-S path 2 had an initially promising free energy profile, up until the user accidentally clashed the ligand against a hydrophobic surface residue. This makes H-S path 2 is of limited value, although, with a small adjustment, the user could instead explore a surface groove near this residue (which would lead the system towards a previously observed metastable state)(30), leading to a lower barrier. The feasibility of such an approach will depend on the size of the system, available hardware and efficiency of the simulation software used. For moderately large protein-ligand complexes, modern high-end GPUs can provide up to microseconds of MD sampling per day for systems of this size, (38, 39) or approximately 10 ps of sampling in 1 second of GPU time. Therefore, an on-the-fly adaptation of this protocol is theoretically within reach.

To minimize computational load and simplify the experimental pipeline, our simulation protocol included some approximations. The iMD-VR simulations here employed implicit solvation, and so for consistency between iMD-VR and US, the same approach was applied throughout. Water molecules are thought to play an important role in benzamidine binding to trypsin, (40) including water-mediated stabilizing interactions for intermediate states, (36) so lack of explicit solvent is not ideal. We also employed protein backbone restraints to limit the conformational space accessible during iMD-VR, to avoid large changes due to the high forces applied. Additional benefits of this are that consistency between H-S paths is improved, because the global protein structure is prevented from diverging between paths, and it allows for faster convergence for both US and path optimization with the ASM. As the protein movement is restricted, some resolution of the unbinding process may be lost. For example, just before exiting the binding pocket in H-S path 7, a favourable interaction is briefly formed with Tyr39. However, a previously suggested metastable state has benzamidine sandwiched between this residue and Tyr151. (30) As the protein had a limited range of movement in our simulations, benzamidine could not contact both tyrosine residues simultaneously during sampling. Backbone restraints are also likely to artificially increase free energy differences between bound and unbound states: there is a bias towards the protein conformation in the starting, bound complex, which in turn causes US to overestimate its stability.

Nonetheless, given that iMD-VR trajectories can include energetic artifacts due to the bias applied by the user (such as the energy spike seen in H-S path 2), there is limited benefit to using an expensive sampling protocol as a first pass. We further do not recommend that calculated free energies from our suggested protocol are treated as accurate. Instead, this protocol should be used to evaluate iMD-VR 'guessed' paths relative to one another. Here, it was found to be more favourable to direct benzamidine through a large groove on the trypsin surface and towards the catalytic triad, especially in the direction of metastable states identified in other work. (30) Refinement using the ASM method can further be used to quantify where iMD-VR generated paths can be improved. Here, the user guided benzamidine from the S1 pocket benzyl group first, resulting in a sharp energy barrier. With ASM refinement, however, the trajectory samples benzamidine rotating out of the S1 pocket, as seen in other work. (30, 36) This gives feedback to the iMD-VR user that they should not pull benzamidine out with the hydrophobic ring directly pointing at the solvent. Such changes towards a more energetically feasible pathway could also be obtained by using 'on-the-fly' integration of iMD-VR with pathCV-based enhanced sampling. Promising pathways could be repeatedly sampled in iMD-VR, whilst optimizing them to reach low energy barriers and metastable states. We propose the following software pipeline: (i) The user generates a trajectory of an unbinding pathway using iMD-VR; (ii) individual snapshots are projected onto 6d-space as soon as they are generated; (iii) the new points are added to the pathCV definition and an additional US window is defined and sampled; (iv) free energy data for the unbinding path up to the new point is passed back to the user. We anticipate this 'on-the-fly' feedback would prompt the user to explore more favourable regions of conformational space, making the iMD-VR session more productive. Additionally, only the more promising paths would be chosen for more extensive sampling and detailed analysis, reducing computational load.

The trypsin-benzamidine system is a good proof-of-principle for our workflow, because it is well understood (through extensive simulation with more conventional approaches). The version of pathCVs used here (26) can characterize a large number of interdependent coordinates in an arbitrary number of dimensions. Therefore, by also including internal coordinates from the protein, this workflow could be used for other protein-ligand systems (e.g. where the motion of a lid-like domain often controls ligand release). While it is unlikely that a single researcher generating a single unbinding pathway will perfectly characterize a MFEP, it is apparent that iMD-VR can be used to quickly sample a range of physically reasonable pathways. It is possible to run iMD-VR remotely, hosted on the cloud, which would allow scientists (and non-scientists) from around the world to be recruited to generate a large ensemble of paths. Generated pathways can be 'scored' by iMD-VR-FE, and hence this problem gamified, with researchers aiming to find paths with low barriers (and metastable states). (13) These data can then guide accelerated sampling methods along these paths, allowing intelligent exploration of complex dynamics. In understanding the relative energetics of a bound state, its surrounding metastable regions and feasible unbinding pathways, users can gain insights that aid drug design, complementing more computationally intensive, non-interactive enhanced sampling approaches. (29-32, 36, 40, 41) Given that iMD-VR can also be used to create protein-ligand complexes for which an experimental structure does not exist, (11) the iMD-VR-FE protocol may be especially suited towards areas such as discovery and development of allosteric drugs. Here, we present an initial implementation of this protocol; we anticipate that it will be developed further for multiple different applications.

# 4. Methods

## 4a. iMD-VR for sampling of protein-ligand unbinding pathways

**System setup**

Trypsin was parameterized with the Amber ff14SB forcefield (42), benzamidine was parameterized with the general amber forcefield (GAFF) in Antechamber (43), and the solvent was modelled implicitly using the OBC2 generalized Born model. (44) Prior to using iMD-VR to generate unbinding pathways, the complexed structure, with starting coordinates from PDB ID 1S0R, was minimized and equilibrated. The details of this process are given in Section 2 of the SI Appendix.

**iMD-VR simulations**

A minimized and equilibrated complex of benzamidine bound to the S1 pocket of trypsin was used as the starting point for iMD-VR. An iMD-VR proficient user then proceeded to carefully guide the ligand out of the binding pocket. Harmonic positional restraints were used for the protein backbone atoms CA, N, O, C (10 $kcal \cdot mol^{-1} \cdot Å^{-2}$) and the $Ca^{2+}$ ion (20 $kcal \cdot mol^{-1} \cdot Å^{-2}$). A total of seven different iMD-VR unbinding pathways were generated, each taking a distinct route (shown in Fig. 2). With the provided Narupa simulation files, researchers can set up their own locally hosted simulation environments. We also make a Narupa iMD-VR demo of the trypsin-benzamidine interactive simulations available via cloud infrastructure, which can be launched from app.narupa.xyz. Instructions for connecting to this demo can be found in Section 1 of the SI Appendix.

## 4b. Calculating free energies along protein-ligand unbinding pathways

**Definition of the pathCVs**

The unbinding pathways obtained from iMD-VR were first characterized by 6 simple CVs describing relative orientation of the two species based on 3 reference points on each, as proposed in (33). These reference points for each species (protein and ligand) were chosen such that their geometric centers form approximately equilateral triangles and their positions are not easily affected through thermal fluctuations. Explicit definitions of the CVs are included in Section 3 of the SI Appendix.

After the pathways were projected onto the selected CV space, the pathCVs were used to: (a) define a reaction coordinate (RC) that changes smoothly along the path, and (b) ensure that the simulation system stays in the vicinity of the path. The metric-corrected (26) version of the pathCVs was used to account for different functional forms and couplings (distance, angles and dihedral angles) of the chosen CVs. Definition of the pathCVs is provided in the SI Appendix (Section 3).

**Free-energy calculation**

The free energy profiles along the pathCVs were calculated using US. (20) This consists of running a set of simulations biased to different values of the chosen RC with harmonic biasing potentials and subsequent integration of the obtained sampling to recover the full free energy profile. The same setup and simulation protocol were used for all 7 H-S paths. Details of the protocol can be found in Section 4 of the SI Appendix. Briefly, all the US simulations were performed with a modified version of sander from AmberTools (https://github.com/kzinovjev/string-amber) using the same parameterization and implicit solvent model as used in iMD-VR. 56 US windows were used. The initial structures for US windows were obtained by taking the closest snapshot from the VR pathway and running 1 ps MD, while gradually increasing the force constant from zero to the target value. 1 ns of sampling was acquired for each window during production simulations. The resulting potentials of mean force were integrated using the weighted histogram analysis method (WHAM). (45) The additional analysis carried out on path 7 (Fig. 3) utilized the adaptive string method. (34) Path optimization was performed using an extension of the sander code, which is available on GitHub (https://github.com/kzinovjev/string-amber). Details of the optimization protocol can be found in Section 5 of the SI Appendix.

## Acknowledgements

K.Z. and M.W.vdK. acknowledge support by the Biotechnology and Biological Sciences Research Council (BB/L018756/1 and BB/M026280/1), the Engineering and Physical Sciences Research Council (EP/V011421/1) and the UK Catalysis Hub (EPSRC grant EP/M013219/1). K.Z. also acknowledges the Maria Zambrano contract at the University of Valencia funded by Ministerio de Universidades (BOE-A-2021-6391). H.M.D. thanks the Engineering and Physical Sciences Research Council (EPSRC) for a PhD studentship. H.M.D. and A.J.M. acknowledge support by the Engineering and Physical Sciences Research Council and UK Catalysis Hub (EP/R026939/1, EP/R026815/1, EP/R026645/1, and EP/R027129/1). AJM acknowledges funding from the European Research Council (ERC) under the European Union's Horizon 2020 research and innovation programme (PREDACTED Advanced Grant, Grant agreement No.: 101021207) and from EPSRC for CCP-BioSim (EP/M022609/1). J.B. acknowledges funding from the EPSRC (programme grant EP/P021123/1). DRG acknowledges support from the European Research Council under the European Union's Horizon 2020 research and innovation programe through consolidator grant NANOVR 866559, and also thanks the Axencia Galega de Innovación for funding as an "Investigador Distinguido" through the Oportunius Program. We thank the Advanced Computing Research Centre of the University of Bristol for computational facilities.

## Supporting Information

The SI Appendix contains instructions for connecting to a cloud-hosted instance of a trypsin-benzamidine simulation, additional details on the MD protocols used in this work, a description of the 6-dimensional CV space, expressions for the pathCVs and the protocols used for umbrella sampling and adaptive string method calculations.

## Data Availability

The files necessary for running a standalone simulation of trypsin and benzamidine in Narupa are available at: https://gitlab.com/intangiblerealities/narupa-protocol. Simulation parameters, input files for Narupa, the seven iMD-VR guided trajectories, reference pathways and the free energy profiles for all the runs are available via DOI: 10.5281/zenodo.6659616 , as additional Supporting Information.

# References

1. P. Leinen *et al.*, Virtual reality visual feedback for hand-controlled scanning probe microscopy manipulation of single molecules. *Beilstein J Nanotechnol* **6**, 2148-2153 (2015).
2. P. Leinen *et al.*, Autonomous robotic nanofabrication with reinforcement learning. *Sci Adv* **6** (2020).
3. S. Ferretti, S. Bianchi, G. Frangipane, R. Di Leonardo, A virtual reality interface for the immersive manipulation of live microscopic systems. *Sci Rep* **11**, 7610 (2021).
4. L. J. Kingsley *et al.*, Development of a virtual reality platform for effective communication of structural data in drug discovery. *J Mol Graph Model* **89**, 234-241 (2019).
5. J. Laureanti *et al.*, Visualizing biomolecular electrostatics in virtual reality with UnityMol-APBS. *Protein Sci* **29**, 237-246 (2020).
6. R. J. Garcia-Hernandez, D. Kranzlmuller, NOMAD VR: Multiplatform virtual reality viewer for chemistry simulations. *Comput Phys Commun* **237**, 230-237 (2019).
7. M. B. O'Connor *et al.*, Interactive molecular dynamics in virtual reality from quantum chemistry to drug binding: An open-source multi-person framework. *The Journal of Chemical Physics* **150**, 220901 (2019).
8. R. K. Walters, E. M. Gale, J. Barnoud, D. R. Glowacki, A. J. Mulholland, The emerging potential of interactive virtual reality in drug discovery. *Expert Opin Drug Discov* 10.1080/17460441.2022.2079632, 1-14 (2022).
9. M. O'Connor *et al.*, Sampling molecular conformations and dynamics in a multiuser virtual reality framework. *Science Advances* **4**, eaat2731 (2018).
10. H. M. Deeks *et al.*, Interactive molecular dynamics in virtual reality for accurate flexible protein-ligand docking. *PLOS ONE* **15**, e0228461 (2020).
11. H. M. Deeks, R. K. Walters, J. Barnoud, D. R. Glowacki, A. J. Mulholland, Interactive Molecular Dynamics in Virtual Reality Is an Effective Tool for Flexible Substrate and Inhibitor Docking to the SARS-CoV-2 Main Protease. *J Chem Inf Model* **60**, 5803-5814 (2020).
12. H. T. H. Chan *et al.*, Discovery of SARS-CoV-2 Mpro Peptide Inhibitors from Modelling Substrate and Ligand Binding. *bioRxiv* (2021).
13. R. J. Shannon *et al.*, Exploring human-guided strategies for reaction network exploration: Interactive molecular dynamics in virtual reality as a tool for citizen scientists. *J Chem Phys* **155**, 154106 (2021).
14. S. Amabilino *et al.*, Training Neural Nets To Learn Reactive Potential Energy Surfaces Using Interactive Quantum Chemistry in Virtual Reality. *J Phys Chem A* **123**, 4486-4499 (2019).
15. S. Amabilino, L. A. Bratholm, S. J. Bennie, M. B. O'Connor, D. R. Glowacki, Training atomic neural networks using fragment-based data generated in virtual reality. *J Chem Phys* **153**, 154105 (2020).
16. B.R. Jagger, S.E. Kochanek, S. Haldar, R.E. Amaro, A.J. Mulholland, Multiscale simulation approaches to modeling drug–protein binding. *Curr Opin Struct Biol* **61**, 213-221 (2020).
17. A. Nunes-Alves, D. B. Kokh, R. C. Wade, Recent progress in molecular simulation methods for drug binding kinetics. *Curr Opin Struct Biol* **64**, 126-133 (2020).
18. I. Dierynck *et al.*, Binding kinetics of darunavir to human immunodeficiency virus type 1 protease explain the potent antiviral activity and high genetic barrier. *J Virol* **81**, 13845-13851 (2007).
19. C. F. Shuman, P. O. Markgren, M. Hämäläinen, U. H. Danielson, Elucidation of HIV-1 protease resistance by characterization of interaction kinetics between inhibitors and enzyme variants. *Antiviral Res* **58**, 235-242 (2003).
20. J. Kästner, Umbrella sampling. *Wiley Interdisciplinary Reviews: Computational Molecular Science* **1**, 932-942 (2011).
21. A. Barducci, M. Bonomi, M. Parrinello, Metadynamics. *Wiley Interdisciplinary Reviews: Computational Molecular Science* **1**, 826-843 (2011).
22. F. Noe, C. Clementi, Collective variables for the study of long-time kinetics from molecular trajectories: theory and methods. *Curr Opin Struct Biol* **43**, 141-147 (2017).
23. K. Zinovjev, S. Marti, I. Tunon, A Collective Coordinate to Obtain Free Energy Profiles for Complex Reactions in Condensed Phases. *J Chem Theory Comput* **8**, 1795-1801 (2012).
24. M. Bonomi, D. Branduardi, F. L. Gervasio, M. Parrinello, The unfolded ensemble and folding mechanism of the C-terminal GB1 beta-hairpin. *J Am Chem Soc* **130**, 13938-13944 (2008).
25. L. Hovan, F. Comitani, F. L. Gervasio, Defining an Optimal Metric for the Path Collective Variables. *J Chem Theory Comput* **15**, 25-32 (2019).

26. K. Zinovjev, I. Tunon, Exploring chemical reactivity of complex systems with path-based coordinates: role of the distance metric. *J Comput Chem* **35**, 1672-1681 (2014).
27. K. Zinovjev, I. Tuñón, Reaction coordinates and transition states in enzymatic catalysis. *WIREs Computational Molecular Science* **8**, e1329 (2018).
28. E. Formoso, V. Limongelli, M. Parrinello, Energetics and Structural Characterization of the large-scale Functional Motion of Adenylate Kinase. *Scientific Reports* **5**, 8425 (2015).
29. L. W. Votapka, B. R. Jagger, A. L. Heyneman, R. E. Amaro, SEEKR: Simulation Enabled Estimation of Kinetic Rates, A Computational Tool to Estimate Molecular Kinetics and Its Application to Trypsin-Benzamidine Binding. *J Phys Chem B* **121**, 3597-3606 (2017).
30. I. Buch, T. Giorgino, G. De Fabritiis, Complete reconstruction of an enzyme-inhibitor binding process by molecular dynamics simulations. *Proc Natl Acad Sci U S A* **108**, 10184-10189 (2011).
31. N. Plattner, F. Noe, Protein conformational plasticity and complex ligand-binding kinetics explored by atomistic simulations and Markov models. *Nat Commun* **6**, 7653 (2015).
32. F. Noe, C. Clementi, Kinetic distance and kinetic maps from molecular dynamics simulation. *J Chem Theory Comput* **11**, 5002-5011 (2015).
33. D. Suh, S. Jo, W. Jiang, C. Chipot, B. Roux, String Method for Protein-Protein Binding Free-Energy Calculations. *J Chem Theory Comput* **15**, 5829-5844 (2019).
34. K. Zinovjev, I. Tuñón, Adaptive Finite Temperature String Method in Collective Variables. *The Journal of Physical Chemistry A* **121**, 9764-9772 (2017).
35. M. Mares-Guia, E. Shaw, Studies on the Active Center of Trypsin. The Binding of Amidines and Guanidines as Models of the Substrate Side Chain. *J Biol Chem* **240**, 1579-1585 (1965).
36. P. Tiwary, V. Limongelli, M. Salvalaglio, M. Parrinello, Kinetics of protein-ligand unbinding: Predicting pathways, rates, and rate-limiting steps. *Proc Natl Acad Sci U S A* **112**, E386-391 (2015).
37. D. Branduardi, F. L. Gervasio, M. Parrinello, From A to B in free energy space. *J Chem Phys* **126**, 054103 (2007).
38. *OpenMM Benchmarks*. Available at: https://openmm.org/benchmarks [Accessed July 14, 2022].
39. Amber20: pmemd.cuda performance information. Available at: https://ambermd.org/GPUPerformance.php [Accessed July 14, 2022].
40. J. Schiebel *et al.*, Intriguing role of water in protein-ligand binding studied by neutron crystallography on trypsin complexes. *Nat Commun* **9**, 3559 (2018).
41. I. Teo, C. G. Mayne, K. Schulten, T. Lelièvre, Adaptive Multilevel Splitting Method for Molecular Dynamics Calculation of Benzamidine-Trypsin Dissociation Time. *J Chem Theory Comput* **12**, 2983-2989 (2016).
42. J. A. Maier *et al.*, ff14SB: Improving the Accuracy of Protein Side Chain and Backbone Parameters from ff99SB. *J Chem Theory Comput* **11**, 3696-3713 (2015).
43. J. Wang, R. M. Wolf, J. W. Caldwell, P. A. Kollman, D. A. Case, Development and testing of a general amber force field. *J Comput Chem* **25**, 1157-1174 (2004).
44. A. Onufriev, D. Bashford, D. A. Case, Exploring protein native states and large-scale conformational changes with a modified generalized born model. *Proteins: Structure, Function, and Bioinformatics* **55**, 383-394 (2004).
45. M. Souaille, B. Roux, Extension to the weighted histogram analysis method: combining umbrella sampling with free energy calculations. *Comput Phys Commun* **135**, 40-57 (2001).

# Free energy along drug-protein binding pathways interactively sampled in virtual reality

## *Supporting Information*

Helen M. Deeks,[1,†] Kirill Zinovjev,[2,3†] Jonathan Barnoud[1,4], Adrian J. Mulholland[1], Marc W. van der Kamp[1,3*], David R. Glowacki,[4*]

[†]*These authors contributed equally this work*
[1]*Center for Computational Chemistry, School of Chemistry, University of Bristol, Bristol, BS8 1TS, UK,*
[2]*Departamento de Química Física, Universidad de Valencia, Burjassot, 46100, Spain,* [3]*School of Biochemistry, University of Bristol, Bristol, BS8 1TD, UK,* [4]*CiTIUS Intelligent Technologies Research Centre, Rúa de Jenaro de la Fuente, s/n, 15705 Santiago de Compostela, A Coruña, Spain*

*\*marc.vanderkamp@bristol.ac.uk; \*drglowacki@gmail.com*

## 1. Connecting to a cloud-hosted iMD-VR simulation

### 1.1 Requirements

The VR client has been developed on Microsoft Windows using the HTC Vive Pro, the Valve Index, the Oculus Rift, the Oculus Rift S, and both the Oculus Quest and Quest 2 using Oculus link. In all cases, a VR-capable computer running both Windows 10 and SteamVR is necessary. The server runs remotely and data is sent to the VR client over the internet. Therefore, a stable internet connection is required. For a simulation size comparable to trypsin-benzamidine, a 10 Mbps download speed is the recommended minimum.

**1.2 Run a server in the cloud**

A Narupa server runs the molecular dynamics simulation, sends the frames to the clients, and synchronises the positions of the user's avatars. The Narupa Cloud interface can be accessed at [https://app.narupa.xyz](https://app.narupa.xyz). Upon first use of the Narupa Cloud service, a user needs to create an account. Only one user needs to start a server instance; once the server is running, other users can connect to it without an account on the service.

Once logged into Narupa Cloud, go to the "Sessions" section of the site. The page lists the current and future server sessions that are scheduled by the user. By clicking on the "Schedule a session" button, a user can start a server session now, or schedule one in the future.

On the scheduling page, select the simulation named "Trypsin-Benzamidine", the starting time and the duration you want. Ideally, you would select the server closest to your location (this improves latency). Clicking the "Schedule" button at the bottom of the page will validate your choices and submit the scheduling request. The scheduled session should then appear in the list on the "Sessions" page. Server sessions scheduled to start immediately needs a few minutes to be ready; session scheduled for the future should be ready at the requested time. When the session is ready, an IP address appears next to the session on the "Sessions" page. This IP address is needed for the VR clients to connect and should be sent to all intended users.

**1.3 Connect a VR client**

Before running the VR client, make sure the VR headset is plugged and ready to go. You need SteamVR installed. More information on SteamVR can be found at (https://store.steampowered.com/app/250820/SteamVR)

Each user needs to download the latest version of Narupa iMD at (https://gitlab.com/intangiblerealities/narupa-applications/narupa-imd/-/jobs/artifacts/master/download?job=build-StandaloneWindows64). The file behind that link is named "artifact.zip", decompress it at the location of your choice. In the resulting directory, move to the "Builds" directory, then the "StandaloneWindows64" one, and double click on the "Narupa iMD.exe" file. Windows will display a security warning because it does not know about Narupa, click on "More info", then "Run anyway". At this point, Narupa iMD should run and should be displayed both on screen and in the VR headset. On the top left of the computer screen, click on "Direct connect", set the address to the IP address of your server session and click "Connect".

**1.4 Adjust the visual representations**

Changing how the molecules are represented in VR requires to connect to the server session with a python client. To install the python client of Narupa, follow the instructions at (https://gitlab.com/intangiblerealities/narupa-protocol#quick-installation-for-a-user)

You will also need to install jupyter to run the provided notebook. In the Anaconda Powershell Prompt used in the previous steps, execute the following commands:

```
conda activate Narupa
conda install jupyter
jupyter notebook /the/path/to/the/downloaded/notebook
```

An example visual representation notebook can be found in the 'simulations' folder of the supplementary materials. Change the IP address at the top of the notebook to the one of the server session and run all the cells.

## 2. Trypsin-benzamidine system set-up

Hydrogen atoms were added to the protein and ligand using reduce. The protein structure was parameterized with the Amber 14 force field (1) and benzamidine was parameterized with the General Amber Forcefield (2) in antechamber, using AM1-BCC partial charges. The solvent in each simulation was modelled implicitly using OBC2. (3)

Prior to simulation in iMD-VR, the structure was minimized and equilibrated in implicit solvent. First, the structure was iteratively energy minimized using slowly decreasing degrees of positional restraint. 5 kcal/mol/$A^2$, 2.5 kcal/mol/$A^2$, and 1.25 kcal/mol/$A^2$ was applied to all backbone and ligand atoms for the first three rounds of minimization respectively, and no restraints were applied for the final round. Next, the system was heated by running 10 stages total of 20 ps of molecular dynamics, starting at 0K and linearly increasing the temperature by 30K at each stage until a temperature of 298K was reached (each step had a backbone and ligand atom restraint of 5 kcal/mol/$A^2$).

Finally, 8 rounds of 500 ps of molecular dynamics with slowly decreasing backbone and ligand atom restraints was run to equilibrate the structure. Restraints were initially 5 kcal/mol/Å² and halved after each step; once backbone restraints were below 1 kcal/mol/A², the restraint atoms were reduced to only C-alpha backbone atoms and ligand atoms. The eighth and final stage had no restraints on the protein and ligand at all. All stages of minimization had a 20 kcal/mol/ A² positional restraints on the calcium ion embedded within the trypsin structure.

Due to the introduction of artificially high forces during iMD-VR, a strong backbone positional restraint was applied to the trypsin structure and single calcium ion during interactive simulations. All other parameters were taken directly from the minimization and test production MD. For all iMD-VR simulations, a temperature of 300K was used with a timestep of 0.5 fs. Snapshots of the iMD-VR simulations were taken every 500 timesteps, equal to every 0.25 ps.

## 3. Definition of the reaction coordinate

The unbinding pathways obtained from VR were characterized by 6 collective variables (CVs) describing relative orientation of the two species as proposed in (4):

$$r = distance(P_1L_1); \theta = angle(P_1L_1L_2); \phi = dihedral(P_1L_1L_2L_3)$$

$$\Theta = angle(P_2P_1L_1); \Phi = dihedral(P_2P_1L_1L_2); \Psi = dihedral(P_3P_2P_1L_1)$$

Where P1-P3 and L1-L3 are reference centers in the protein and the ligand respectively, defined as geometric centers of the following atom groups:

| Protein centers (all non-hydrogen atoms of following residues) | Ligand centers (atoms) |
|---|---|
| P1: Cys136, Leu137, Ile138, Cys157, Leu158, Lys159, Pro198, Val199, Val200 | L1: C, C1, N1, N2 |
| P2: Phe181, Cys182, Ala183, Val213, Ser214, Trp215, Gly226, Val227, Tyr228 | L2: C2, C3 |
| P3: Val31, Ser32, Leu33, Phe41, Cys42, Gly43, Gln64, Val65, Arg66 | L3: C5, C6 |

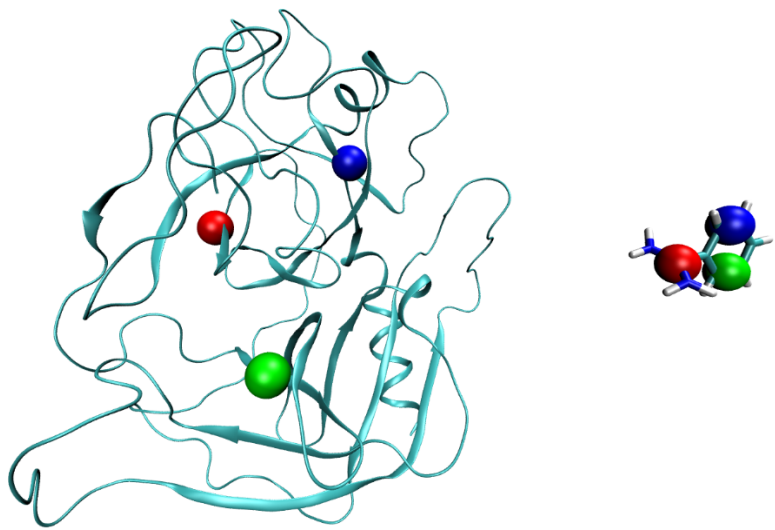

**Figure S1: Geometric centers used in CV definition.** P1/L1, P2/L2/ and P3/L3 are represented with red, green and blue spheres respectively.

These reference points for each species (protein and ligand) were chosen such that their geometric centers form approximately equilateral triangles that are not sensitive to thermal fluctuations, which in case of the protein means avoiding the use of residues present in flexible loops. By choosing this representation we assume that the internal degrees of freedom (such as vibrations and conformational changes) are not coupled to the unbinding and therefore can be excluded from the reaction coordinate. While this assumption is valid for the trypsin-benzamidine case, it would fail if there were a significant conformational rearrangement during the unbinding (e.g. lid opening-like motion). In such case, an additional CV describing the conformational change must be included.

To obtain the free energy profiles along iMD-VR pathways one needs to 1) define a reaction coordinate (RC) that changes smoothly along the path and 2) ensure that the simulation system stays in the vicinity of the path. The latter is especially important when the user intentionally generates a sub-optimal path to explore an alternative unbinding mechanism. Without any additional restraint, there is a high chance of the system reverting to a more favorable path (which might have already been explored). Both tasks can be accomplished with path collective variables (pathCVs) (5) defined along the paths in the chosen CV space (6):

$$s(r) = \frac{\sum_{i=1}^{n} t_i e^{-\lambda d(\theta(r), z(t_i))}}{\sum_{i=1}^{n} e^{-\lambda d(\theta(r), z(t_i))}}; \quad z(r) = -\lambda^{-1} \ln \sum_{i=1}^{n} e^{-\lambda d(\theta(r), z(t_i))}; \quad t_i = \frac{i-1}{n-1} L$$

Where $z(t)$ is the reference path parameterized by the arc length, $L$ is the total length of the path and $d(\theta(r), z)$ is some distance measure between the given state of the system $r$ and a point $z$ on the reference path and $\theta(r)$ is the 6-vector of CVs defined above. $\lambda$ is set to inverse of the distance between the points $z(t_i)$. To account for interdependency and different nature of the CVs, the following measure was used (7):

$$d(\theta(r), z(t_i)) = \sqrt{(\theta(r) - z(t_i))^T M(t_i)^{-1} (\theta(r) - z(t_i))}$$

Where $M$ is the variable distance metric tensor defined as:

$$M(t_i)_{jk} = \langle \nabla \theta_j(r) \cdot \nabla \theta_k(r) \rangle_{\theta(r) = z(t_i)}$$

Where gradients are taken with regards to mass-weighted Cartesian coordinates and $\langle ... \rangle_{\theta(r) = z(t_i)}$ denotes canonical ensemble average over configurations constrained to the point $z(t_i)$ in the CV space.

## 4. Umbrella sampling protocols

The same setup and simulation protocol were used for all 7 human-sampled paths. All the simulations were performed with a modified version of sander from AmberTools19 (https://ambermd.org/CiteAmber.php). Generalized Born implicit solvation method was used to describe the water solution. Temperature was set to 300K and was controlled with Langevin thermostat. The bonds involving hydrogen were constrained using SHAKE, which allowed to set the integration timestep to 2fs. 56 Umbrella Sampling windows were used. Harmonic biases were equally spaced along the range of pathCV values with force constants determined automatically to guarantee uniform sampling assuming flat underlying free energy profile (see (8) for details). The initial structures for US windows were obtained by taking the closest snapshot from the VR pathway and running 1 ps MD gradually increasing the force constant from 0 to the target value. 1 ns of sampling was acquired during production simulations. Hamiltonian replica exchange between windows was attempted every 500 fs. To restrain the sampling to the vicinity of the path, a harmonic bias was added along the $z$ coordinate at $z = 0$ with force constant = $1\ kcal \cdot mol^{-1} \cdot a.m.u.^{-1} \cdot Å^{-2}$. The resulting potentials of mean force were integrated using WHAM procedure. (9)

## 5. String method protocols

The adaptive version (8) of on-the-fly string method (10) was used to obtain the minimum free energy path in the vicinity of path 7. The following ASM parameters were changed from their default values to ensure faster convergence and stability of the simulations: $\gamma = 500\ ps^{-1}, K^{\perp} = 1\ kcal \cdot mol^{-1} \cdot a.m.u.^{-1} \cdot Å^{-2}, \gamma_{\alpha} = 200\ kcal \cdot mol^{-1} \cdot a.m.u.^{-1} \cdot Å^{-2} \cdot ps, \kappa = 1\ a.m.u.^{-2} \cdot Å^{-4}$. Same CVs as for the pathCV Umbrella Sampling calculations were used in the string method. Path 7 was used as the initial guess. Endpoints of the path were fixed in the CV space. The MD protocol was identical to the one used for Umbrella Sampling. The string optimization converged after 1.7 ns of simulation. The free energy along the converged path was obtained using the same procedure as for the paths obtained directly from VR.

## References

1. Maier JA, Martinez C, Kasavajhala K, Wickstrom L, Hauser KE, Simmerling C. ff14SB: Improving the Accuracy of Protein Side Chain and Backbone Parameters from ff99SB. J Chem Theory Comput. 2015;11(8):3696-713.
2. Wang J, Wolf RM, Caldwell JW, Kollman PA, Case DA. Development and testing of a general amber force field. J Comput Chem. 2004;25(9):1157-74.
3. Onufriev A, Bashford D, Case DA. Exploring protein native states and large-scale conformational changes with a modified generalized born model. Proteins: Structure, Function, and Bioinformatics. 2004;55(2):383-94.
4. Suh D, Jo S, Jiang W, Chipot C, Roux B. String Method for Protein-Protein Binding Free-Energy Calculations. J Chem Theory Comput. 2019;15(11):5829-44.
5. Branduardi D, Gervasio FL, Parrinello M. From A to B in free energy space. J Chem Phys. 2007;126(5):054103.

6. Zinovjev K, Marti S, Tunon I. A Collective Coordinate to Obtain Free Energy Profiles for Complex Reactions in Condensed Phases. J Chem Theory Comput. 2012;8(5):1795-801.

7. Zinovjev K, Tunon I. Exploring chemical reactivity of complex systems with path-based coordinates: role of the distance metric. J Comput Chem. 2014;35(23):1672-81.

8. Zinovjev K, Tuñón I. Adaptive Finite Temperature String Method in Collective Variables. The Journal of Physical Chemistry A. 2017;121(51):9764-72.

9. Kumar S, Rosenberg JM, Bouzida D, Swendsen RH, Kollman PA. THE weighted histogram analysis method for free-energy calculations on biomolecules. I. The method. Journal of Computational Chemistry. 1992;13(8):1011-21.

10. Maragliano L, Vanden-Eijnden E. On-the-fly string method for minimum free energy paths calculation. Chemical Physics Letters. 2007;446(1):182-90.